\documentstyle[11pt,psfig]{article}
\newcommand{\RR}{\rangle \rangle}
\newcommand{\LL}{\langle \langle}
\begin{document}
\renewcommand{\thepage}{ }
\begin{titlepage}
\title{
{\center \bf  Antiferromagnetism
in a doped spin-Peierls model: classical and
quantum behaviors}}
\author{
R. M\'elin\thanks{melin@labs.polycnrs-gre.fr}
{}\\
{Centre de Recherches sur les Tr\`es Basses
Temp\'eratures (CRTBT)\thanks{U.P.R. 5001 du CNRS,
Laboratoire conventionn\'e avec l'Universit\'e Joseph Fourier
}}\\
{CNRS BP 166X, 38042 Grenoble Cedex, France}\\
{}\\
}
\date{\today}
\maketitle
\vspace{2cm}
\begin{abstract}
\normalsize
We address the problem of
antiferromagnetism in a two dimensional
model of doped spin-Peierls
system, at the classical and quantum levels. 
A Bethe-Peierls solution is derived for the
classical model, with an ordering temperature
proportional to the doping concentration.
The quantum model is treated
in a cluster renormalization group showing
a finite randomness behavior and an antiferromagnetic
susceptibility at low temperature.
\end{abstract}
\end{titlepage}
\newpage
\renewcommand{\thepage}{\arabic{page}}
\setcounter{page}{1}
\baselineskip=17pt plus 0.2pt minus 0.1pt

\newpage

\section{Introduction}

The spin-Peierls transition at
$T_{\rm SP} \simeq 14$~K
in the inorganic quasi one dimensional
spin-Peierls compound CuGeO$_3$ has attracted
much interest~\cite{SP}. This transition is
characterized by the appearance of a 
finite dimerization in the CuO$_2$ chains,
and the opening of a spin gap.
With CuGeO$_3$, it became possible to
experiment the effect of doping in a spin-Peierls
system. 
An antiferromagnetic (AF) 
phase was discovered upon replacing a fraction of the
Cu ions (with $S=1/2$) by
magnetic ions with a different spin: Ni~\cite{Ni} (with $S=1$)
or Co~\cite{Co} (with $S=3/2$), or non magnetic ions:
Zn~\cite{Zn,Hase,Martin} or Mg~\cite{Mg}.
Also, the Ge sites (outside the CuO$_2$ chains)
can be substituted with Si~\cite{Si},
leading to antiferromagnetism at low
temperature.

Recent experiments by Manabe {\sl et al.}
have shown
the existence of a finite N\'eel temperature
$\simeq 25$~mK,
with a doping concentration
as low as $0.12$~$\%$~\cite{Manabe98}.
The doping dependence of the N\'eel
temperature obtained in these experiments
suggests the absence
of a critical concentration for the appearance
of antiferromagnetism: at low doping the 
ordering temperature scales like
$\ln{T_N} \propto 1 /x$~\cite{Manabe98}.

Early theoretical works~\cite{Martins96,Fukuyama,Khomskii}
have focussed on the identification of the relevant
low energy degrees of freedom. Fukuyama, Tanimoto and
Saito~\cite{Fukuyama} have shown
the coexistence between 
antiferromagnetism and dimerization
in a doped spin-Peierls model.
The degrees of freedom relevant to
the low energy physics are
solitonic spin-$1/2$ excitations pinned
at the impurities~\cite{Khomskii}.
These excitations are the building
blocks of the theory in Refs.~\cite{Fabrizio97a,Fabrizio97b,Fabrizio99}.
These spin-1/2 objects interact {\sl via} an exchange
decaying exponentially with distance. Interchain interactions
can be incorporated by considering
the existence of a transverse correlation length,
approximately one tenth of the
longitudinal correlation length, as recently
proposed independently by
Dobry {\sl et al.}~\cite{Dobry99}, and
Fabrizio, M\'elin and Souletie~\cite{Fabrizio99}.
Numerical calculations with realistic
spin-phonon couplings have provided a link
between the microscopic Hamiltonian
and the effective model of interacting
spin-$1/2$
moments~\cite{Martins96,Laukamp98,Hansen98a,Hansen98b,Augier98}.
The approach followed in
Refs.~\cite{Fabrizio97a,Fabrizio97b,Fabrizio99}
and continued in the present work
relies on the treatment of 
disorder in the effective Hamiltonian.
This allows to discuss the 
qualitative physics of large scale systems
at a finite temperature,
while the numerical methods have so far been limited
to the ground state
properties~\cite{Dobry99,Martins96,Laukamp98,Hansen98a,Hansen98b,Augier98}.
The scope of this article is to analyze the
model beyond the percolation approximation
used in Ref.~\cite{Fabrizio99}, both at
the classical and quantum levels.
We first
show in section~\ref{sec:MF}
that the physics of the quantum Hamiltonian
is already present in the classical Ising
Hamiltonian, and  give a rigorous derivation
of mean field theory {\sl via} a Bethe-Peierls
treatment. The second purpose of the article
is to show that the quantum Hamiltonian
has an antiferromagnetic behavior at low
temperature. The
quantum model is treated in a cluster
renormalization group (RG) calculation
in sections~\ref{sec:qu-iso} and~\ref{sec:qu-ani}.

\section{The model}
\label{sec:model}
We recall the model proposed in
Ref.~\cite{Fabrizio97a}.
When 
impurities are introduced in a dimerized
background (for instance non magnetic
impurities such as Zn), spin-1/2 solitonic moments
are released out of the dimerized pattern.
These magnetic moments are
pinned at the impurities due to interchain
interactions~\cite{Khomskii}.
This picture is in agreement with susceptibility
experiments~\cite{Manabe98,Grenier,Saint-Paul},
indicating the release of one
spin-$1/2$ moment per Zn impurity at low doping. 
The interaction between two magnetic moments
at a distance $d$ originates from virtual excitations
of the gaped dimerized background, and decays
exponentially with distance, with a characteristic
length set by the correlation lengths
$\xi_x \sim 9$~c along the chain direction
(c-axis)~\cite{Kiryukin,Horvatic},
and $\xi_y \sim \xi_x /10$ in the b-axis direction.
These exchange
interactions as well as the relevance of disorder
were identified in Ref.~\cite{Fabrizio97a}
to play a crucial role in the establishment of
antiferromagnetism. 
The low energy physics of a doped spin-Peierls
system is represented by
spin-$1/2$ solitonic magnetic moments
distributed at random with a concentration $x$
on a square lattice,
and interacting {\sl via} a Heisenberg
Hamiltonian~\cite{Fabrizio99,Dobry99}
\begin{equation}
\label{eq:H-1}
H = \sum_{\langle i,j \rangle}
J_{i-j} {\bf S}_i . {\bf S}_j
,
\end{equation}
the exchange in Eq.~\ref{eq:H-1} being staggered
and decaying exponentially
with distance:
\begin{equation}
\label{eq:H-2}
J_{i-j} = (-1)^{d_x + d_y + 1}
\Delta \exp{\left( - \sqrt{
\left( \frac{d_x}{\xi_x} \right)^2
+ \left( \frac{d_y}{\xi_y} \right)^2
}\right)}
,
\end{equation}
with $\xi_x\simeq 9$~c the correlation length
along the c-axis and $\xi_y \simeq 0.1 \times \xi_x$
the correlation length along the 
b-axis. Correlations along the a-axis
are neglected.

\section{Classical Ising model}
\label{sec:MF}
We consider a model with
Ising degrees of freedom, distributed randomly and
interacting {\sl via} the exchange in Eq.~\ref{eq:H-2}.
The
classical antiferromagnet has the same transition
temperature as the classical ferromagnet.
We consider therefore the 
ferromagnetic model to
calculate the ordering temperature.

\subsection{One dimensional model}
\label{sec:1D}

\begin{figure}
\vspace*{1cm}
\centerline{\psfig{file=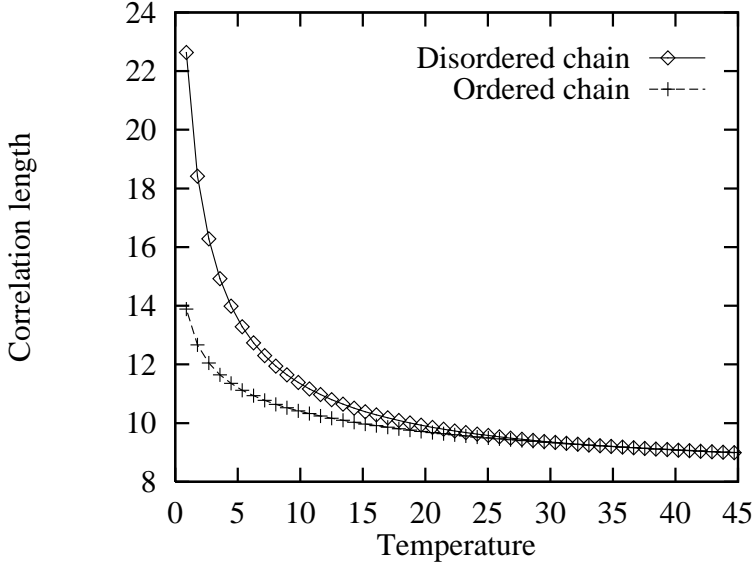,height=7.5cm}}
\caption{The average correlation length of the
disordered Ising
spin chain with nearest neighbor couplings
is larger than the
correlation length of an ordered system with the same
concentration $x=0.01$ above $T^*$.
We used $\Delta=44.7$~K, $\xi=9$.
}
\label{fig:corre1D}
\end{figure}

A high temperature expansion leads to
the exact form of 
the correlations in terms of a product over
the bonds between the spins at sites $0$
and $L$:
$
\langle \sigma_0 \sigma_L \rangle
= \prod \tanh{(\beta J_{i,i+1})}
.
$
We calculated numerically the disorder average
to obtain the correlation length at a finite
temperature. As shown on Fig.~\ref{fig:corre1D},
the average correlation length of the disordered model
is larger than the typical correlation length
$\xi = -1 / [x \ln{(\tanh{(\beta T^*)})}]$,
with $T^* = \Delta
\exp{(-1/(x \xi))}$ the exchange of
the particular
disorder realization where the magnetic
moments are equally spaced. 
We calculated in Ref.~\cite{Fabrizio97b}
the correlation length of the quantum
chain and found a similar result:
the enhancement of the magnetic correlations
above $T^*$
due to disorder does not rely on the quantum
nature of the coupling Hamiltonian in spite
of a random singlet physics in the quantum
chain, not present in the Ising chain.

\subsection{Determination of the exchange distribution}
\label{sec:PJ}
We consider the
exchanges to be drawn independently in a 
distribution $P(J)$
resulting from the combination of
randomness in the spatial distribution of the
magnetic moments and
exponentially decaying interactions, Eq.~\ref{eq:H-2}.
The relevant exchanges are set by
the spins the closest to each other. Therefore,
given a spin at site $(x_0,y_0)$, we need to
determine the probability that one spin is found
on the periphery of the ellipse $ [(x-x_0) / \xi_x]^2
+ [ (y-y_0) / \xi_y ]^2 = \gamma^2$,
with no other spin inside the ellipse,
and therefore an exchange $\Delta \exp{(
- \gamma)}$.
We consider a system of total
area $A$ containing $n$ spins.
The probability to find no spin inside a
subsystem of area $\delta A$ is
$
P_0 = \left( 1 - \frac{ \delta A}{A} \right)^n
\simeq \exp{(- x \delta A)}
$,
with $x = n/A$ the doping concentration.
Now the spacing distribution is
\begin{equation}
\label{eq:P-lambda}
P(\gamma) = x L(\gamma) \exp{(-x \delta A(\gamma))}
,
\end{equation}
with $L(\gamma) = d [A(\gamma)] / d \gamma$.
In the one dimensional model, we have
$\delta A(\gamma)= 2 \gamma \xi_x$, and
$L(\gamma)=2 \xi_x$.
In the two dimensional isotropic model with
$\xi_x = \xi_y = \xi$,
we have
$\delta A(\gamma) = \pi \gamma^2 \xi^2$, and
$L(\gamma) = 2 \pi \gamma \xi$.
In the quasi one dimensional model,
$\delta A(\gamma) = \pi
\gamma^2 \xi_x \xi_y$, and
$L(\gamma) = 2 \pi \gamma
\xi_x \xi_y$.
The distribution $P(\gamma)$ of the isotropic 
and anisotropic two dimensional models
is a Wigner distribution with a short scale
``distance repulsion''.
This repulsion will be shown not to affect the
ordering properties.

\subsection{Bethe-Peierls transition in the
infinite coordination limit}
\label{sec:infinite-z}
\begin{figure}
\vspace*{1cm}
\centerline{\psfig{file=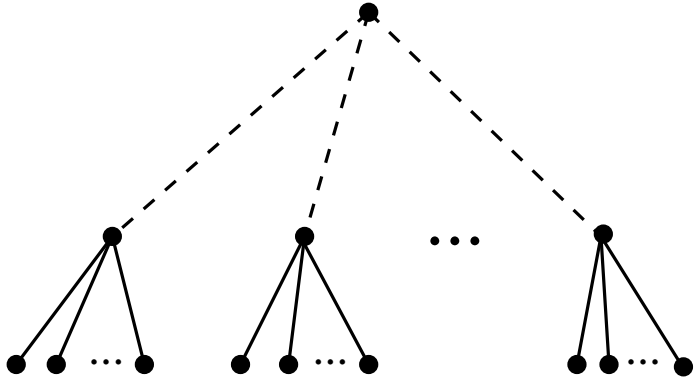,height=5cm}}
\caption{The tree structure used in the Bethe-Peierls
calculation. The forward branching ratio is $z-1$.
A tree with $n+1$ generations is obtained from
connecting $z-1$ trees with $n$ generations.
}
\label{fig:BP}
\end{figure}

We now consider the Bethe-Peierls solution of
the Ising model~\cite{Baxter}. 
The lattice has a tree topology,
with a forward branching ratio $z-1$, and
we calculate the magnetization of the
site with the highest hierarchical level
(top spin),
in the presence of the other sites.
We consider $z-1$ trees and
connect them to obtain a tree
with one more generation
(see Fig.~\ref{fig:BP}).
The recursion of the average magnetization
of the top spin reads~\cite{Carlson90}
\begin{equation}
\label{eq:X}
X = \frac{ \prod_{i=1}^{z-1} ( 1 + Y_i
\tanh{(\beta J_i)}) -
\prod_{i=1}^{z-1} (1 - Y_i \tanh{(\beta J_i)})}
{ \prod_{i=1}^{z-1} ( 1 + Y_i
\tanh{(\beta J_i)}) +
\prod_{i=1}^{z-1} (1 - Y_i \tanh{(\beta J_i)})}
,
\end{equation}
where $Y_i$, $i=1,...,z-1$ are the magnetizations of
the top spins with $n$ generations, and 
$X$ the
magnetization of the top spin with $n+1$
generations.
We first consider the artificial situation where
the ordering temperature is large compared
to the exchange: $T_{\rm bp} \gg \Delta$, which
turns out to be equivalent to 
assuming a large coordination.
Eq.~\ref{eq:X} is linearized into
$
X = \sum_{i=1}^z Y_i \tanh{(\beta J_i)}
$,
leading to the recursion of the magnetization
$
\LL X \RR_{n+1} = (z-1) \LL \tanh{(\beta J)} \RR
\LL X \RR_n
$, with the subscript $n$ labeling the number of
generations. This leads to the
ordering temperature
$
T_{\rm bp} = (z-1) \LL J \RR
$,
far above $\Delta$ if $z \gg 1$, and consistent
with the initial assumption.
We can calculate the ordering
temperature $T_{\rm bp}$ with the different
distributions $P(J)$ derived in section~\ref{sec:PJ}.
We find:
\begin{itemize}
\item[(i)] {\sl with the one dimensional model distribution}:
$T_{\rm bp} = 2 (z-1) x \xi \Delta / (1 + 2 x \xi)$;

\item[(ii)] {\sl with the isotropic two dimensional model
distribution}:
$T_{\rm bp} \simeq 2 (z-1) \pi x \xi^2 \Delta$ in
the dilute regime $x \xi^2 < 1$.
\item[(iii)] {\sl with the quasi one dimensional model
distribution}:
$T_{\rm bp} \simeq 2 (z-1) \pi x \xi_x \xi_y \Delta$
in the dilute regime $x \xi_x \xi_y <1$.
\end{itemize}
The three limits therefore show a similar
behavior $T_{\rm bp} \propto (z-1) x \Delta$,
showing that the short distance Wigner
repulsion in the spacing distribution
Eq.~\ref{eq:P-lambda}
does not affect the ordering properties.
Comparing the Bethe-Peierls ordering temperature
to the ordering temperature obtained from the
Stoner criterion in Ref.~\cite{Fabrizio99},
we see that $z-1$ should be identified with
the interchain coupling. The small-$z$ regime,
relevant to weak interchain correlations,
is now discussed in sections~\ref{sec:BP-perco}
and~\ref{sec:beyond-perco}.

\subsection{Bethe-Peierls transition
with a finite coordination: (i) percolation
approximation}
\label{sec:BP-perco}
We now consider the physics at a finite
$z=3$. In this regime, the Bethe-Peierls
method takes into account
the inhomogeneities of the magnetization,
not included in the Stoner criterion
mean field solution
in Ref.~\cite{Fabrizio99}.
We first consider a
``percolation approximation'' in which we
assume 
the bonds $J \ll T$ ($J \gg T$) to be set to zero
(infinity) in the effective percolation problem.
With $z=3$, the Bethe-Peierls iteration Eq.~\ref{eq:X}
reads
$$
X = \frac{ Y \tanh{(\beta J_y)}
+ Z \tanh{(\beta J_z)}}
{1 + YZ \tanh{(\beta J_y)}
\tanh{(\beta J_z)}}
,
$$
and is approximated into:
(i) $T \ll J_y$, $T \ll J_z$:
$X \simeq Y + Z$;
(ii) $T \gg J_y$,
$T \ll J_z$: $X \simeq Z$;
(iii) $T \ll J_y$,
$T \gg J_z$: $X \simeq Y$;
(iv) $T \gg J_y$,
$T \gg J_z$: $X \simeq 0$.
The recursion of the average
magnetization is therefore
$
\LL X \RR_{n+1} \simeq 2 \lambda
\LL X \RR_n
$,
with the percolation parameter
\begin{equation}
\label{eq:lambda}
\lambda = \int_{T}^{+ \infty} P(J) dJ
.
\end{equation}
With the one dimensional distribution,
we have $\lambda = 1 - (T / \Delta)^{2 x \xi_x}$,
which yields a transition at the temperature
\begin{equation}
\label{eq:T*}
T^* = \Delta \exp{\left( - \frac{2 \ln{2}}{x \xi_x}
\right)},
\end{equation}
exponentially
small in $1/(x \xi_x)$.
This behavior 
is compatible with Ref.~\cite{Fabrizio99}
where we have shown the absence of a true ordering
transition in the percolation approximation
of a two dimensional anisotropic model,
while the model was shown to percolate in a
finite size.

\subsection{Bethe-Peierls transition 
with a finite coordination: (ii) beyond the
percolation approximation}
\label{sec:beyond-perco}
We now solve the Ising model 
beyond the percolation
approximation. We take into account
the iteration of small exchanges to lowest
order, with the following approximate
iteration:
(i) $T \ll J_y$, $T \ll J_z$:
$X \simeq Y + Z$;
(ii) $T \gg J_y$,
$T \ll J_z$: $X \simeq \beta J_y Y + Z$;
(iii) $T \ll J_y$,
$T \gg J_z$: $X \simeq Y + \beta J_z Z$;
(iv) $T \gg J_y$,
$T \gg J_z$: $X \simeq \beta J_y Y + \beta J_z Z$.
The dominant contribution  originates from
the region (iv) of the couplings:
$
\LL X \RR_{n+1} \simeq (2/T) \mu (1- \lambda)
\LL X \RR_n
$,
with $\lambda$ in Eq.~\ref{eq:lambda}, and
$
\mu = \int_0^T J P(J) d J
$.
With the one dimensional distribution for $P(J)$, 
we have $\mu 
\simeq 2 x \xi \Delta$ and therefore
the same critical temperature
$T_{\rm bp} = 4 x \xi \Delta$ as 
in the model with
a large connectivity $z$ (with $z=3$ in
this calculation).
It is remarkable that the correct
treatment of the small exchanges
restores a transition temperature 
$\propto x \xi \Delta$. 
This shows the relevant
role played by energy scales
smaller than the temperature.

The main unsolved question regarding the
classical model behavior is to determine
whether the finite dimensional model 
has a true thermodynamic transition at
a temperature $\propto x \xi \Delta$.
The Bethe-Peierls solution orders at
a temperature $\propto x \xi \Delta$
because of the strong short range correlations.
This does not necessarily mean that the
finite dimensional model also has a true
thermodynamic transition at this temperature.
Instead, we believe it possible that the classical model
has a cross-over to a Griffiths physics at a temperature
$\propto x \xi \Delta$ and a true thermodynamic
transition with a diverging correlation length
at a temperature $T^*$, which would also be a behavior
compatible with a low temperature antiferromagnetic
susceptibility. At the present stage, we cannot make
the distinction between these two behaviors.

\section{Quantum isotropic model}
\label{sec:qu-iso}
\begin{figure}
\vspace*{1cm}
\centerline{\psfig{file=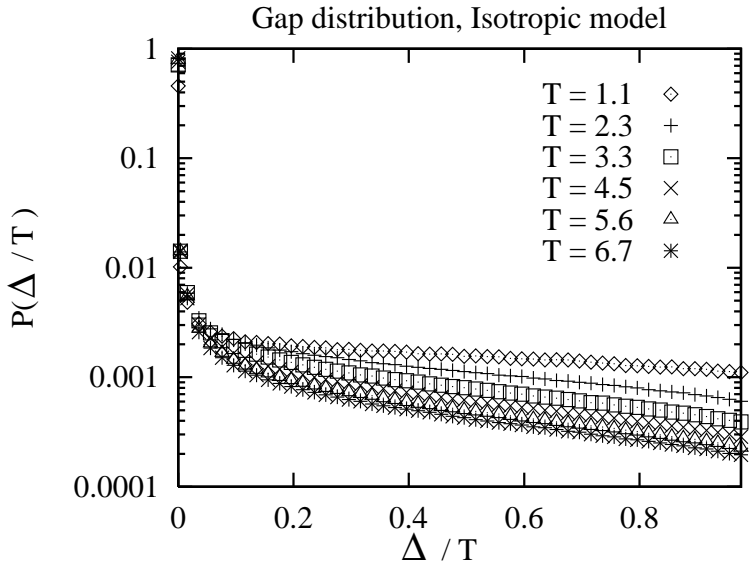,height=7.5cm}}
\caption{Evolution of the gap distribution
of the model with isotropic exchanges
$\xi_x = \xi_y =9$, $\alpha=2$
as the temperature is scaled down.
The doping concentration is $x=0.01$
and the system has
a size $200 \times 200$.
The weight of energy scales
$\sim T$ increases as the temperature
is decreased.
}
\label{fig:P-gap-iso}
\end{figure}

\begin{figure}
\vspace*{1cm}
\centerline{\psfig{file=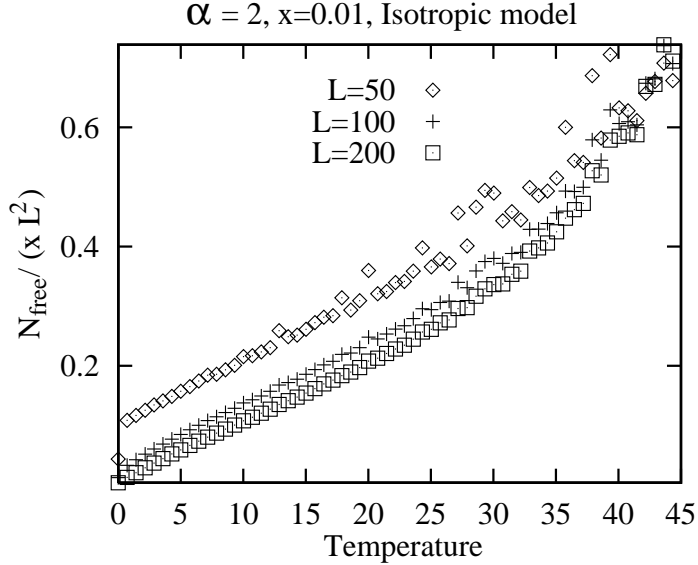,height=7.5cm}}
\caption{Temperature dependence of the number
of effective moments, normalized to the number of
initial moments $x L^2$.
We have set $\xi_x=\xi_y=9$, $\Delta = 44.7$~K,
$x=0.01$, $\alpha=2$. We used square systems of
dimensions $L \times L$, with
$L=50$ ($\Diamond$), $L=100$ ($+$),
and $L=200$ ($\Box$). For large system sizes,
$N_{\rm eff} \sim (T / \Delta) x L^2$.
}
\label{fig:N-eff-iso}
\end{figure}

\begin{figure}
\vspace*{1cm}
\centerline{\psfig{file=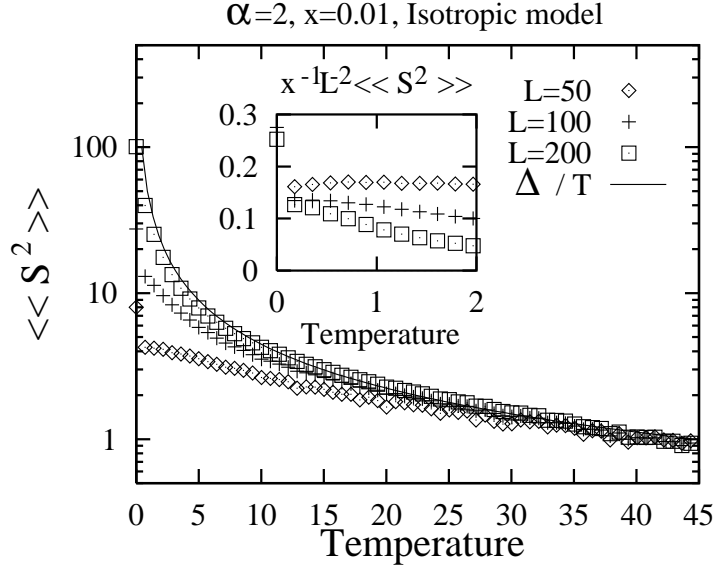,height=7.5cm}}
\caption{Temperature dependence of the
the average square effective moment
$\LL [S^{\rm eff}]^2 \RR$
with $\xi_x=
\xi_y =9$, $\Delta = 44.7$~K, $x=0.01$,
$\alpha=2$. We used square systems
of dimensions $L \times L$, with $L=50$
($\Diamond$), $L=100$ ($+$), $L=200$ ($\Box$).
The solid line is $\Delta / T$.
The insert shows the low 
temperature dependence of
$(x L^2)^{-1} \LL [S^{\rm eff}]^2 \RR$.
}
\label{fig:S2-iso}
\end{figure}

\begin{figure}
\vspace*{1cm}
\centerline{\psfig{file=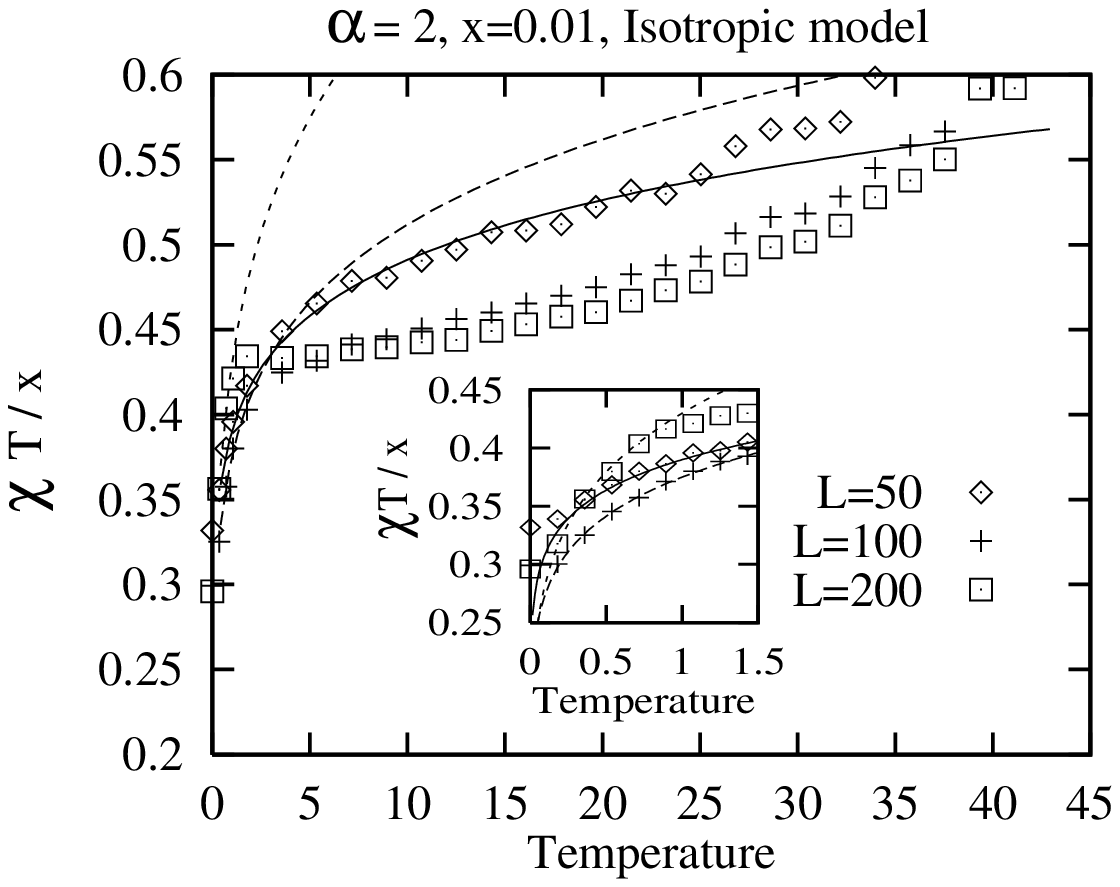,height=7.cm}}
\caption{Temperature dependence of
the Curie constant
$T \chi(T)/x$  with $\xi_x=
\xi_y =9$, $\Delta = 44.7$~K,
and for $\alpha=2$. We used
square systems
of dimension $L \times L$, with $L=50$
($\Diamond$), $L=100$ ($+$), $L=200$ ($\Box$).
The low temperature regime has been fit to a power
law behavior $\chi T/x = 0.39 \times T^{-0.1}$
($L=50$, solid line),
$\chi T/x = 0.375 \times T^{0.135}$ ($L=100$, long dashed line),
and $\chi T/x = 0.43 \times T^{0.18}$ ($L=200$, short dashed line).
The insert shows the low temperature behavior.
}
\label{fig:Curie-iso}
\end{figure}

We first consider the artificial situation where the correlation
lengths are identical in the two directions:
$\xi_x = \xi_y = \xi = 9$. 
The tendency to ordering in this isotropic model
is overestimated compared
to the anisotropic model with $\xi_x=9$,
$\xi_y=0.1 \times \xi_x$. 
We are lead to consider 
the class of interactions
\begin{equation}
\label{eq:H-gene}
J_{i-j} = (-1)^{d_x + d_y + 1}
\Delta \exp{\left( - \left[
\left( \frac{d_x}{\xi_x} \right)^2
+ \left( \frac{d_y}{\xi_y} \right)^2
\right]^{\alpha/2} \right)}
,
\end{equation}
decaying faster than the interactions in
Eq.~\ref{eq:H-2} if $\alpha>1$.
The cluster RG (see the Appendix) generates
large energy scales
in the parameter range
$\alpha < \alpha_{0}\simeq 1.2$.  It turns out that 
$\alpha_0 < 1$ in the model
with anisotropic exchanges, and we therefore
consider only the regime
$\alpha > \alpha_{0} \simeq 1.2$
in the isotropic model.

The gap distribution is shown on Fig.~\ref{fig:P-gap-iso}
for decreasing temperatures. It is visible that
the RG produces gaps of
order of the temperature $T$
unlike in the case of
the infinite randomness fixed
point  where the
opposite occurs (see Ref.~\cite{Fisher94}
for the one dimensional Heisenberg chain
with an infinite randomness, random singlet
behavior; see Ref.~\cite{Motrunich99}
for the infinite randomness behavior in the
two dimensional Ising model in a transverse
field).
As in the
Ising model analysis, we calculate
the susceptibility in two ways:
(i) we assume a paramagnetic behavior of
the set of effective moments;
(ii) we incorporate the correlations
induced by exchanges $\Delta \sim T$,
in which case an antiferromagnetic
behavior in the susceptibility is restored.

\subsection{Infinite randomness calculation}
\label{sec:infinite-random}
We first consider all the exchanges
$J < T$ to be set to zero:
the set of effective spins is viewed as
a paramagnet with a susceptibility
\begin{equation}
\label{eq:sus1}
\chi = \frac{1}{TL^2}  \LL \sum_{i=1}^{N^{({\rm eff})}}
S^{({\rm eff})}_i ( S^{({\rm eff})}_i +1) \RR
,
\end{equation}
where  $N^{({\rm eff})}$ the number of effective spins.
We have discarded a prefactor $1/3$ in Eq.~\ref{eq:sus1},
not relevant to the present calculation.
The low temperature susceptibility is therefore controlled
by two quantities: (i) the density of free spins
$n_{\rm eff} =
\LL N^{({\rm eff})}\RR / (x L^2)$; and
(ii) the magnitude of the effective spin.

The number of effective moments
scales like $N_{\rm eff} \sim (T / \Delta) x L^2 $,
as it is visible on
Fig.~\ref{fig:N-eff-iso}. The squared effective
moment shows two regimes:
\begin{itemize}
\item[(i)] {\sl High temperature regime}: 
The high temperature
average squared effective moment scales like
$\LL [S^{\rm eff}]^2 \RR \sim \Delta / T$
(see Fig.~\ref{fig:S2-iso}).
The susceptibility per unit volume is
$\chi \sim x/T$.

\item[(ii)] {\sl Low temperature percolation regime}:
At low temperature, the squared effective moment
scales like $\LL S^2 \RR \sim a x L^2$, with $a$ some
constant (see the insert Fig.~\ref{fig:S2-iso}).
The susceptibility per unit volume is
$\chi \sim (a x^2 L^2)/\Delta$.
In this regime, a cluster has percolated
through the finite size system. Its magnetization
results from summing $x L^2$ variables
$S_i^z = \pm 1/2$, corresponding to the
two sublattices magnetizations. Therefore,
$\LL S \RR \sim \sqrt{x L^2}$, and
$\LL S^2 \RR \sim x L^2$.
\end{itemize}
The cross-over between these regimes occurs
at the temperature scale $T_{\rm co} =
(\Delta/a) (x L^2)^{-1}$, which decreases to zero when
the system size is increased. Therefore in the
thermodynamic limit, only the high temperature
paramagnetic behavior survives while in a finite
size, 
a low temperature tail is present in the susceptibility
(see Fig.~\ref{fig:Curie-iso}).
Now the situation changes when correlations
between spins coupled by exchanges of order $T$
are included.

\subsection{Finite randomness calculation}
\label{sec:finite-random}

\begin{figure}
\vspace*{1cm}
\centerline{\psfig{file=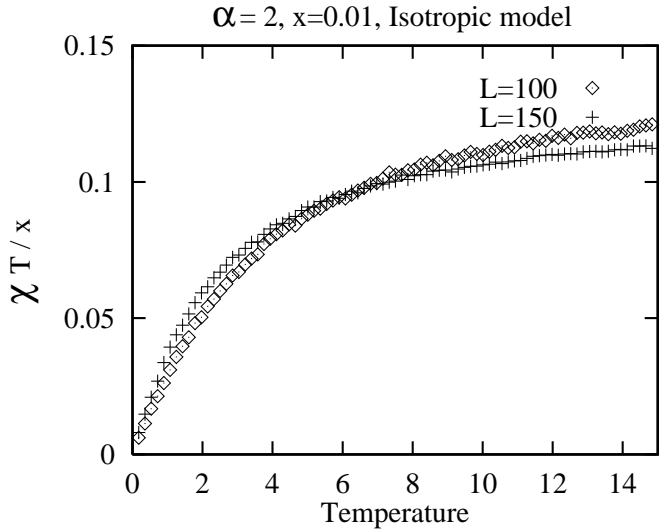,height=7.cm}}
\caption{Temperature dependence of the Curie
constant $\chi T/x$ with an isotropic system of size
$L \times L$, with $L=100$, $L=150$.
The spins connected by exchanges between
$T/2$ and $T$ have been frozen. This results
in a finite Curie constant at zero temperature.
}
\label{fig:chi-iso}
\end{figure}

To schematically incorporate the
correlations at
energy scales of order of the temperature,
we consider as frozen the spins connected
by an exchange with a gap between $T/2$ and $T$.
This freezing results in a staggered magnetization
because the set of effective moments is unfrustrated
(see the Appendix).
The resulting
susceptibility is shown on Fig.~\ref{fig:chi-iso}.
It is visible that
$\chi T$ is linear in $T$ at small
$T$, with therefore
a finite susceptibility at zero temperature.
This shows qualitatively how an antiferromagnetic
behavior can be restored
because of the correlations at energy
scales $\Delta \sim T$.

\section{Quantum anisotropic model}
\label{sec:qu-ani}
\subsection{One dimensional model}
In one dimension, the RG equations of a model in
which only AF nearest neighbor 
exchanges are retained can be solved exactly
(see Ref.~\cite{Fabrizio97b}).
We note $x = J / [\mbox{Max}(J)]$ the
exchange normalized to the maximal exchange. The
distribution of the variable $x$ is
$P(x) = (\overline{f}/ \Gamma) x^{ \overline{f}/\Gamma
-1}$, with $\Gamma = \ln{(\Delta / \mbox{Max}(J))}$, and
$\overline{f} / \Gamma = x \xi / ( 1 + \Gamma x \xi)$.
The weight on the strongest exchanges
$x \simeq 1$ is $\simeq x \xi$ above the cross-over
temperature $T^* = \Delta \exp{(-1/ x \xi)}$,
and decreases to zero at temperatures
below $T^*$, where the system has crossed over
to the random singlet fixed point.
As a test of our program, we considered the cluster
RG of a one dimensional model, with
the exchanges not restricted to nearest
neighbors (see Eqs.~\ref{eq:H-1},~\ref{eq:H-2}).
For any practical temperature above $T^*$,
the weight on energy scales of order $T$
is found to remain constant as the system
is renormalized,
with therefore the same behavior
as in the one dimensional model
with AF nearest neighbor exchanges only.

\subsection{Anisotropic model}

\begin{figure}
\vspace*{1cm}
\centerline{\psfig{file=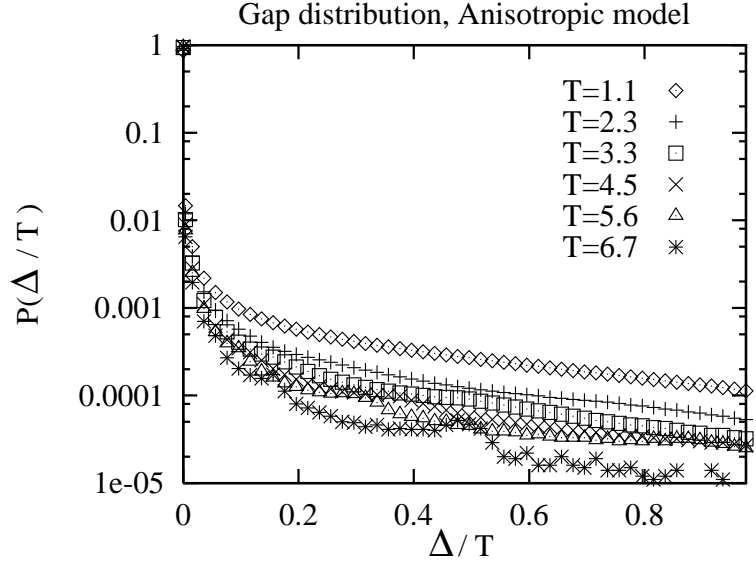,height=7.5cm}}
\caption{Gap distribution of the
anisotropic model, with $\xi_x=9$, $\xi_y=0.1 \times \xi_x$,
and $x=0.01$ and a system of
size $L_x \times L_y$, with
$L_x=640$ and $L_y=64$. The oscillatory behavior
at high temperature is due to the anisotropy
in the couplings.
}
\label{fig:ani-gap}
\end{figure}
\begin{figure}
\vspace*{1cm}
\centerline{\psfig{file=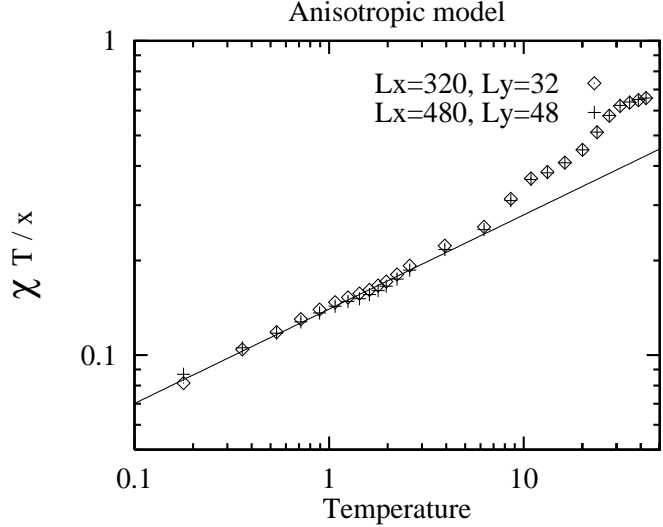,height=7.cm}}
\caption{Temperature dependence of the Curie
constant $\chi T / x$ of the anisotropic model,
with $\xi_x = 9$,  $\xi_y = 0.1 \times \xi_x$,
and $x=0.01$. The lattice sizes are
$320 \times 32$ and
$480 \times 48$. The solid line is a fit
of the low temperature behavior to the
form $\chi T = 0.14 \times T^{0.3}$.
}
\label{fig:chi-ani}
\end{figure}

We show on Fig.~\ref{fig:ani-gap} the evolution of the
gap distribution as the system is scaled down,
with the parameters $\xi_x=9$, $\xi_y=0.1 \times \xi_x$,
relevant to CuGeO$_3$.
As in
the isotropic model, energy scales of order $T$
are generated upon renormalizing the system.
To qualitatively include the effects of correlations
at energy scales $\Delta \sim T$,
we consider the spins connected by
an exchange with a gap between $T/2$ and $T$
to be frozen, and obtain a low temperature
power-law Curie susceptibility, $\chi \sim
T^{\alpha}$, with
$\alpha =- 0.7$ if $x=0.01$ (see Fig.~\ref{fig:chi-ani}).
The low temperature susceptibility
diverges slower than a Curie law,
which is a behavior characteristic of an
antiferromagnet.
We did not succeed to obtain
$\alpha >0$ as it is the case in doped
CuGeO$_3$. Therefore, we cannot rigorously conclude
on whether antiferromagnetism is long ranged
or associated to a zero temperature transition.
A precise discussion of this point
is an open question, and would require
the correlations at energy $\Delta \sim T$
to be incorporated beyhond our present treatment.
For instance the cluster RG could be used to
renormalize the high energy physics and the
low energy effective Hamitonian could be treated
by exact diagonalizations.

\section{Conclusions}
We have shown that the physics of the quantum
Hamiltonian Eqs.~\ref{eq:H-1},~\ref{eq:H-2}
was already present at the level of the classical
Ising model. A Bethe-Peierls treatment of the
classical model has been given in which a
transition at a temperature $\propto x \xi \Delta$
was found. The quantum Hamiltonian has been treated
in a cluster RG. The model was shown to have a 
finite randomness behavior. We have shown
at a qualitative level how a low temperature
antiferromagnetic susceptibility can be obtained.

Two questions are left open:
\begin{itemize}
\item[(i)] The Bethe-Peierls
solution orders at a temperature $\propto x \xi \Delta$.
We do not know whether the two dimensional model
has also a thermodynamic transition at a temperature
$\propto x \xi \Delta$, or whether this temperature
scale corresponds to a cross-over to a Griffith
physics. Both behaviors would be {\sl a priori}
compatible with the existence of a maximum in
the susceptibility of the antiferromagnet
at a temperature $\propto x \xi \Delta$.

\item[(ii)] The quantum model susceptibility shows an
antiferromagnetic behavior at low temperature
due to correlations at energies $\Delta \sim T$.
The isotropic model shows a finite susceptibility at
low temperature while the quasi one dimensional
has a susceptibility diverging slower than
a Curie law. A precise investigation of the
low temperature susceptibility would require
a treatment going beyond our present analysis,
for instance by treating the low energy
effective Hamiltonian by exact diagonalizations.

\end{itemize}

Two other proposals to explain antiferromagnetism in
doped in CuGeO$_3$ have been made: Fukuyama,
Tanimoto and Saito~\cite{Fukuyama} and
Mostovoy, Khomskii, and J. Knoester~\cite{Mostovoy97}.
These proposals are quite different from
ours, and we have exposed previously why
we think they our model is more relevant~\cite{Fabrizio99}.
The inclusion of interchain interaction 
in our model in Ref.~\cite{Fabrizio99} and
the present work,
points strongly towards an compatibility with
experiments.

\section*{Acknowledgements}
The author thank M. Fabrizio and J. Souletie for
numerous fruitful discussions, their encouragements,
and for useful comments on the manuscript.
M. Fabrizio pointed out to me the possibility
of cluster RG calculations as well as the proof
that the effective problem remains unfrustrated
as the system is scaled down. J. Souletie suggested
the existence of a similar physics in the classical
and quantum models. The cluster RG calculations
have been performed on the CRAY T3E supercomputer
of the Centre Grenoblois de Calcul Vectoriel of
the Commisariat \`a l'Energie Atomique.


\appendix

\begin{figure}
\vspace*{1cm}
\centerline{\psfig{file=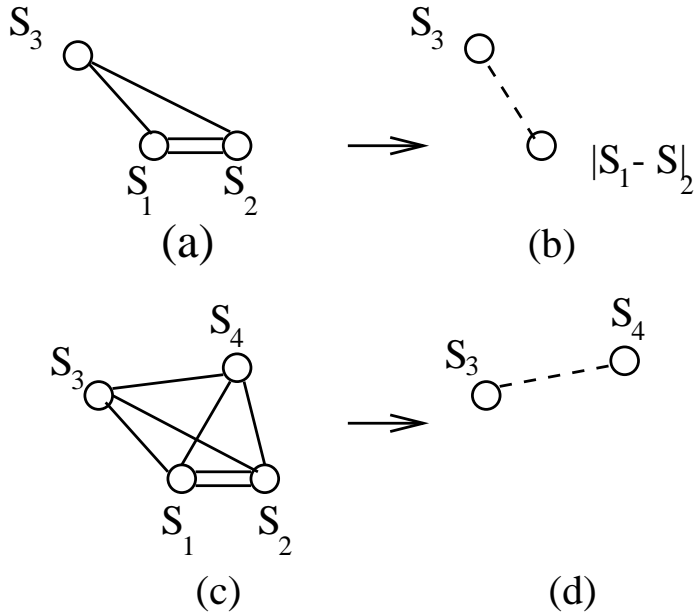,height=8cm}}
\caption{The first RG transformations in a cluster
expansion with a residual spin formation (a) and
(b), and a singlet formation (c) and (d).
The dashed lines represent
renormalized exchanges.
}
\label{fig:schema}
\end{figure}

\section{Renormalization equations}
\label{app:RG}
We use a cluster RG to renormalize
the quantum Hamiltonian
Eqs.~\ref{eq:H-1},~\ref{eq:H-2}. The method
relies on a perturbative expansion
in the inverse of the largest
exchange, and
was originally proposed by Dasgupta and
Ma~\cite{Dasgupta80} in the context
of disordered Heisenberg chains.
The cluster RG
was applied by Bhatt and Lee~\cite{Bhatt81}
to a model of phosphorus doped silicon.
Fisher used the method
to solve exactly the random
singlet fixed point~\cite{Fisher94}.
The cluster RG was also used to investigate the low
energy physics of disordered
spin chains: the dimerized Heisenberg
chain with random exchanges~\cite{Hyman96};
the spin-one chain with random
exchanges~\cite{Hyman97,Monthus97};
Heisenberg chains with 
random ferromagnetic and antiferromagnetic
couplings~\cite{Westerberg96}.
Recently, Motrunich {\sl et al.}~\cite{Motrunich99}
shown the existence of an
infinite randomness
fixed point in two  dimensions
in the Ising model in a transverse field.
At such a fixed point,
inhomogeneities in the disorder
grow indefinitely as the system is
scaled down, as in the random singlet
fixed point in one spatial dimension.
We now give a short derivation of the
RG equations.

We isolate two spins ${\bf S}_1$
and ${\bf S}_2$ coupled by an exchange
$J_{1-2}$. This sets an energy scale
given by the gap between the ground state
and the first excited multiplet:
if $J_{1-2}>0$ is antiferromagnetic,
the ground state has a spin $S=|S_1 - S_2|$
and the first excited multiplet has
$S=|S_1-S_2|+1$, with a gap
$\Delta_{1-2} = |J_{1-2}| (|S_1 - S_2| +1)$.
If $J_{1-2}<0$ is ferromagnetic,
the ground state has $S=S_1+S_2$
and the first excitated multiplet
has $S=S_1+S_2-1$, with a gap
$\Delta_{1-2} = |J_{1-2}| (S_1 + S_2)$.
Among all possible pairs of spins, we consider
the one with the strongest gap $\Delta_{1-2}$.
This energy scale is identified to the system
temperature. 
If ${\bf S}_1$ and ${\bf S}_2$
are coupled ferro ($J_{1-2}<0$) ,
the two spins ${\bf S}_1$ and
${\bf S}_2$ are replaced
by an effective spin $S=S_1 + S_2$.
If they are coupled antiferro, they are replaced by
an effective spin $S=|S_1-S_2|$.
$S_1=S_2$ with an AF coupling $J_{1-2}$
leads to singlet formation while a residual
moment is formed otherwise.

\subsection{Residual moment formation}

Let us first consider the case where
a residual moment is formed corresponding
to (a) and (b) on Fig.~\ref{fig:schema}.
We specialize a spin ${\bf S}_3$
among the other spins and denote
by $J_{i-j}$ the exchange between spins $i$
and $j$, with $i,j=1,...,3$.
The coupling Hamiltonian
between the spins ${\bf S}_1$ and
${\bf S}_2$ is
$
H_{1-2} = J_{1-2} {\bf S}_1 . {\bf S}_2
$
while the remaining couplings
\begin{equation}
\label{eq:HI}
H_I = J_{1-3} {\bf S}_1 . {\bf S}_3
+ J_{2-3} {\bf S}_2 . {\bf S}_3
\end{equation}
are treated in a first order perturbation.
This leads to the renormalized coupling Hamiltonian
$H_I = \tilde{J}_3 {\bf S}_3 . {\bf S}
$, with
the renormalized exchange
\begin{equation}
\label{eq:J-c}
\tilde{J}_3 = J_{1-3} c(S_1,S_2,S) 
+ J_{2-3} c(S_2,S_1,S)
,
\end{equation}
with
$$c(S_1,S_2,S) = \frac{ S(S+1) + S_1 (S_1+1)
- S_2 ( S_2+1)}{2S(S+1)}
$$
derived in Ref.~\cite{Westerberg96}.
The sublattice on which the
residual spin is placed is determined
as follows: if $S_1> S_2$, the residual
spin $S$ is placed on the same
sublattice as $S_1$ while it
is placed on the sublattice of
$S_2$ if $S_1 < S_2$.

\subsection{Singlet formation}

We now consider singlet formation,
with $S_1 = S_2$ coupled AF.
The renormalized couplings are obtained in a
second order perturbation theory.
Generalizing the calculation in
Ref.~\cite{Bhatt81,Westerberg96} to the coupling
Hamiltonian (\ref{eq:HI}), we find
the renormalized exchange
\begin{equation}
\label{eq:Jtilde}
\tilde{J}_{3-4} = J_{3-4}
+ \frac{2 S_1 ( S_1+1)}{3 J_{1-2}}
(J_{1-3} - J_{2-3})
(J_{2-4} - J_{1-4})
,
\end{equation}
where $S_1=S_2$ denote the spins
at site $1$ and $2$. In the 1D limit
$J_{2-3}=J_{1-4}=0$ Eq.~\ref{eq:Jtilde}
reproduces the result in Ref.~\cite{Westerberg96},
and the spin-1/2 limit $S_1=1/2$ reproduces
the result in Ref.~\cite{Bhatt81}.

\subsection{Absence of frustration}
We show that frustration is not generated by the
RG procedure. We assume an unfrustrated starting
Hamiltonian, and show that the different RG operations
are compatible with the sublattice structure. We distinguish
three cases:
\begin{itemize}
\item[(i)] {\sl $S_1$ and $S_2$ belong to
different sublattices and are coupled antiferro}.
We assume $S_1 > S_2$ and the effective spin
$S=S_1 - S_2$ replaces the spin $S_1$. The
renormalized coupling to another spin $S_3$
in Eq.~\ref{eq:J-c} is
$$
\tilde{J}_3 = J_{1-3} + (J_{1-3} - J_{2-3})
\frac{S_2}{S+1}
.
$$
$J_{1-3}>0$ and $J_{2-3}<0$ leads to
$\tilde{J}_3>0$. $J_{1-3}<0$
and $J_{2-3}>0$ leads to $\tilde{J}_3<0$.
The renormalized coupling $\tilde{J}_3$
has thus a sign compatible with the sublattice
structure.

\item[(ii)] {\sl $S_1$ and $S_2$ belong to
the same sublattices and are coupled ferro}.
$J_{1-3}$ and $J_{2-3}$ have the same sign.
We have $c(S_1,S_2,S)>0$ and $c(S_2,S_1,S)>0$.
The renormalized coupling
$\tilde{J}_3$ has the same sign as $J_{1-3}$ and $J_{2-3}$,
compatible with the sublattice structure.

\item[(iii)] {\sl $S_1 = S_2$ are coupled antiferro
and a singlet is formed}. If $S_3$ and $S_4$
are coupled ferro and in the same sublattice,
$\tilde{J}_{3-4}<0$ in Eq.~\ref{eq:Jtilde}.
If  $S_3$ and $S_4$
are coupled antiferro and in the opposite sublattice,
$\tilde{J}_{3-4}>0$ in Eq.~\ref{eq:Jtilde}.
The singlet formation is thus compatible with
the sublattice structure.

\end{itemize}


\newpage

\begin{thebibliography}{99}

\bibitem{SP} M. Hase, I. Terasaki and K.
Uchinokura, Phys. Rev. Lett. {\bf 70}, 3651 (1993);
J.P. Pouget, L.P. Regnault, M. Ain, B. Hennion,
J.P. Renard, P. Veillet, G. Dhalenne
and A. Revcolevschi,
Phys. Rev. Lett. {\bf 72}, 4037 (1994).

\bibitem{Ni} J.-G. Lussier, S.M. Coad, D.F. McMorrow and
D. McK Paul, J. Phys. Condens. Matter {\bf 7},
L325 (1995).

\bibitem{Co} P.E. Anderson, J.Z. Liu and R.N. Shelton,
Phys. Rev. B {\bf 56} (1997), 11014. 

\bibitem{Zn} M. Hase, N. Koide, K. Manabe, Y. Sasago,
K. Uchinokura and A. Sawa, Physica B {\bf 215}, 164 (1995).

\bibitem{Hase} M. Hase, K. Uchinokura, R.J. Birgeneau,
K. Hirota and G. Shirane, J. Phys. Soc. Jpn. {\bf 65}, 1392 (1996).

\bibitem{Martin} M.C. Martin, M. Hase,
K. Hirota and G. Shirane, Phys.
Rev. B {\bf 56}, 3173 (1997).

\bibitem{Mg} T. Masuda, A. Fujioka, Y. Uchiyama, I. Tsukada, and 
K. Uchinokura, Phys. Rev. Lett. {\bf 80}, 4566 (1998).

\bibitem{Si} J.-P. Renard, K. Le Dang, P. Veillet,
G. Dhalenne, A. Revcolevschi and L.P. Regnault,
 Europhys. Lett. {\bf 30}, 475 (1995);
L.P. Regnault, J.P. Renard, G. Dhalenne and
A. Revcolevschi, Europhys. Lett. {\bf 32},
579 (1995).

\bibitem{Manabe98} K. Manabe, H. Ishimoto, N. Koide, Y. Sasago, and 
K. Uchinokura, Phys. Rev. B {\bf 58}, R575 (1998).

\bibitem{Martins96} G.B. Martins, E. Dagotto,
J.A. Riera, Phys. Rev. B {\bf 54}, 16032 (1996).

\bibitem{Fukuyama} H. Fukuyama, T. Tanimoto, and M. Saito, J. Phys. Soc. Jpn. 
{\bf 65}, 1182, 1996. 

\bibitem{Khomskii} D. Khomskii, W. Geertsma and M.
Mostovoy, Czech. Journ. of Phys. {\bf 46} (1996) Suppl S6,
LT21 Conference Proceedings.

\bibitem{Fabrizio97a} M. Fabrizio and R. M\'elin,
Phys. Rev. Lett. {\bf 78}, 3382 (1997).

\bibitem{Fabrizio97b} M. Fabrizio and R. M\'elin,
Phys. Rev. B {\bf 56}, 5996 (1997).

\bibitem{Fabrizio99} M. Fabrizio, R. M\'elin,
and J. Souletie, Eur. Phys. J. B {\bf 10},
607 (1999).

\bibitem{Dobry99} A. Dobry, P. Hansen, J. Riera, D. Augier 
and D. Poilblanc, Phys. Rev. B {\bf 60}, 4065 (1999).

\bibitem{Laukamp98} M. Laukamp, G.B. Martins, C. Gazza,
A.L. Malvezzi, E. Dagotto, P.M. Hansen, A.C. L\'opez
and J. Riera, Phys. Rev. B {\bf 57}, 10755 (1998).

\bibitem{Hansen98a} P.M. Hansen, J.A. Riera, A. Delia and
E. Dagotto, Phys. Rev. B {\bf 58}, 6258 (1998).

\bibitem{Hansen98b} P. Hansen, D. Augier, J. Riera and
D. Poilblanc,
Report cond-mat/9805325.

\bibitem{Augier98} D. Augier, P. Hansen, D. Poilblanc, J.
Riera and E. Sorensen, Report No cond-mat/9805386.


\bibitem{Grenier} B. Grenier, J.P. Renard, P. Veillet,
C. Paulsen, R. Calemcsuk, G. Dhalenne and A. Revcolevschi,
Phys. Rev. B {\bf 57} (1998), 3444.

\bibitem{Saint-Paul} M. Saint-Paul, J. Voiron, C. Paulsen,
P. Monceau, G. Dhalenne and A. Revcolevschi,
J. Phys.: Cond. Matt. {\bf 10}, 10215 (1998).

\bibitem{Kiryukin} V. Kiryukin,
B. Keimer, J.P. Hill, and A. Vigliante, Phys. Rev. Lett.
{\bf 76}, 4608 (1996).

\bibitem{Horvatic} M. Horvati\'c, Y. Fagot-Revurat,
C. Berthier, G. Dhalenne, and A. Revcolevschi,
Phys. Rev. Lett. {\bf 83}, 420 (1999).

\bibitem{Baxter} R.J. Baxter, {\sl Exactly solved
models in statistical mechanics},
Academic Press (1982).

\bibitem{Carlson90} J.M. Carlson, J.T. Chayes,
L. Chayes, J.P. Sethna, and D.J. Thouless,
J. Stat. Phys. {\bf 61}, 987 (1990)

\bibitem{Fisher94} D.S. Fisher, Phys. Rev. B
{\bf 50}, 3799 (1994).

\bibitem{Motrunich99} O. Motrunich, S-C Mau,
D.A. Huse, and D.S. Fisher,
Report No cond-mat/9906322.

\bibitem{Mostovoy97} M. Mostovoy, D. Khomskii,
and J. Knoester, Phys. Rev. B {\bf 58}, 8190
(1998).

\bibitem{Dasgupta80} C. Dasgupta and
S.K. Ma, Phys. Rev. B {\bf 22}, 1305 (1980).

\bibitem{Bhatt81} R.N.~Bhatt and P.A.~Lee,
Phys. Rev. Lett. {\bf 48}, 344 (1982).

\bibitem{Hyman96} R. A. Hyman, K. Yang, R.N. Bhatt,
and S.M. Girvin, Phys. Rev. Lett. {\bf 76}, 839 (1996).

\bibitem{Hyman97} R.A. Hyman and K. Yang, 
Phys. Rev. Lett. {\bf 78}, 1783 (1997).

\bibitem{Monthus97} C. Monthus, O. Golinelli, and
Th. Jolicoeur, Phys. Rev. Lett. {\bf 79},
3254 (1997)

\bibitem{Westerberg96} E.~Westerberg, A.~Furusaki,
M.~Sigrist, and P.A.~Lee, Phys. Rev B
{\bf 55}, 12578 (1997).

\end{thebibliography}
\end{document}